\documentclass[aps,prd,twocolumn,showpacs,superscriptaddress,preprintnumbers, floatfix,nofootinbib]{revtex4-1}
\usepackage{graphicx}
\usepackage{amsmath, amsthm, amssymb}
\usepackage{slashed}
\usepackage{url}
\usepackage[pdftex]{hyperref}
\usepackage{color}
\usepackage{bm}
\usepackage{mathtools}
\usepackage[normalem]{ulem}
\usepackage{comment}

\newcommand{\bea}{\begin{eqnarray}}
\newcommand{\eea}{\end{eqnarray}}

\begin{document}
\preprint{KEK-QUP-2024-0022, KEK-TH-2655, KEK-Cosmo-0359}

\title{Phase Separation Unlocks Resonant Leptogenesis}

\author{Jason Arakawa}
\email{arakawaj@udel.edu}
\affiliation{Department of Physics and Astronomy, University of Delaware, Newark, Delaware 19716, USA}
\affiliation{International Center for Quantum-field Measurement Systems for Studies of the Universe and Particles (QUP, WPI),
High Energy Accelerator Research Organization (KEK), Oho 1-1, Tsukuba, Ibaraki 305-0801, Japan}
\affiliation{Theory Center, Institute of Particle and Nuclear Studies (IPNS), High Energy Accelerator Research Organization (KEK), Tsukuba 305-0801, Japan
}

\author{Philip Lu}
\email{philiplu11@gmail.com}
\affiliation{Center for Theoretical Physics, Department of Physics and Astronomy, Seoul National University, Seoul 08826, Korea}
\affiliation{School of Physics, Korean Institute for Advanced Study, Seoul 02455, Korea}

\author{Volodymyr Takhistov}
\email{vtakhist@post.kek.jp}
\affiliation{International Center for Quantum-field Measurement Systems for Studies of the Universe and Particles (QUP, WPI),
High Energy Accelerator Research Organization (KEK), Oho 1-1, Tsukuba, Ibaraki 305-0801, Japan}
\affiliation{Theory Center, Institute of Particle and Nuclear Studies (IPNS), High Energy Accelerator Research Organization (KEK), Tsukuba 305-0801, Japan
}
\affiliation{Graduate University for Advanced Studies (SOKENDAI), \\
1-1 Oho, Tsukuba, Ibaraki 305-0801, Japan}
\affiliation{Kavli Institute for the Physics and Mathematics of the Universe (WPI), UTIAS, \\The University of Tokyo, Kashiwa, Chiba 277-8583, Japan}

\date{\today}
\begin{abstract}
Distinct behavior of decaying particles in true and false vacua can lead to enhanced baryon asymmetry generation, a scenario we call phase separation leptogenesis. We demonstrate that phase separation enables small mass splittings of right-handed neutrinos that can trigger resonant leptogenesis inside long-lived false vacuum remnants at low energy scales, while allowing for hierarchical neutrino masses at present.  The mechanism significantly expands the parameter space for right handed neutrino mass splittings by multiple orders of magnitude viable for realizing resonant leptogenesis by suppressing washout in the true vacuum and eliminating the requirement that degeneracy persists at late cosmological times. We present a concrete realization in a minimal model with a scalar field undergoing a first-order phase transition, highlighting possible connections with neutrino mass hierarchies and gravitational waves.
\end{abstract}

\maketitle 

{\it Introduction.}-- The observed baryon asymmetry of the Universe (BAU) presents a major puzzle that is challenging to address with just the Standard Model (SM) alone. To dynamically generate baryon asymmetry successfully, 
new physics must satisfy several essential conditions~\cite{Sakharov:1967dj}. That is, models of baryogenesis must violate baryon number, violate charge-parity CP symmetry, and exhibit out-of-equilibrium dynamics. 
Such conditions are often achieved in theories through considerations of decays of new beyond SM heavy particles.

An elegant and well-studied minimal framework to address BAU is leptogenesis~\cite{Fukugita:1986hr,Davidson:2008bu}. There, out of equilibrium CP-violating decays of heavy right-handed neutrinos (RHN) generate a lepton asymmetry that is subsequently converted to observed baryon asymmetry through the electroweak sphaleron processes. A lower bound~\cite{Davidson:2002qv} of around $\sim10^{9}$ GeV on the mass of the lightest RHN required to generate BAU in standard thermal leptogenesis presents significant challenges for direct experimental probes. Intriguingly, this is alleviated when RHN masses are nearly degenerate, with the mass splitting being comparable to their decay width,
CP-violating decays of RHNs can be dramatically resonantly enhanced~\cite{Pilaftsis:2003gt,Pilaftsis:2005rv}.
However, achieving such RHN mass-splitting typically requires additional fine-tuning or model complexity~\cite{Pilaftsis:2005rv,BhupalDev:2014pfm}.

In this work, we propose a novel mechanism of baryogenesis based on enhanced generation of baryon asymmetry through the drastically distinct behavior of decaying particles in true and false vacua of the theory. We demonstrate this in the context of leptogenesis, where resonantly-enhanced lepton asymmetry from RHN decays without fine-tuning of their masses can be naturally generated. 
Unlike standard resonant leptogenesis scenarios, our mechanism accommodates a broader range of neutrino mass hierarchies that can still generate lepton asymmetry resonantly. This feature enables additional connections with neutrino observations and applicable to a wide class of underlying theories.

Our mechanism can readily arise during cosmological first-order phase transitions that produce false-vacuum remnants.  Recently, scenarios of dark matter and leptogenesis~\cite{Chway:2019kft,Wong:2023qon,Chao:2020adk,Baker:2019ndr,Huang:2022vkf,Dasgupta:2022isg, PhysRevLett.130.031803,Fernandez-Martinez:2020szk,Fernandez-Martinez:2022stj} exploiting lack of such vacuum remnant formation have been proposed, and related considerations can also result in formation of primordial black holes~\cite{Kawana:2021tde,Kawana:2022olo,Kim:2023ixo} and axion remnants have also been recently considered~\cite{Carenza:2024tmi}. Baryogenesis associated with particles trapped in the false vacuum and relying on their increased annihilation rates has also been considered~\cite{Arakawa:2021wgz}. Ref.~\cite{Huang:2023gse} introduced a scenario with phenomenological dynamical couplings, where RHN decays are suppressed in the early Universe and switched on in the true vacuum. In strongly first-order phase transitions the vacuum bubble walls can accelerate, sweeping RHNs into the true vacuum and generating large induced masses \cite{Dasgupta:2022isg,Huang:2022vkf,Chun:2023ezg}. Both approaches exploit the true vacuum to trigger rapid RHN decays and suppress washout via Boltzmann factors.

In contrast, our scenario explores the complementary regime where RHNs decay inside false-vacuum remnants. For moderate and weakly first-order transitions, the bubble walls slow down, allowing RHNs to accumulate and form semi-stable false-vacuum regions. Washout remains suppressed in the bulk true vacuum, while concentrated RHNs decay in the remnants. Unlike in strong transitions where vacuum energy drives wall acceleration, here the vacuum energy is deposited as latent heat into the false-vacuum regions, reheating the RHNs and raising the dark-sector temperature. This false-vacuum realization of phase separation leptogenesis dynamically triggers resonant enhancement of the baryon asymmetry while suppressing washout, and ultimately yields a hierarchical RHN mass spectrum today.

\begin{figure*}[t]
    \centering
    \includegraphics[width = 0.49\linewidth]{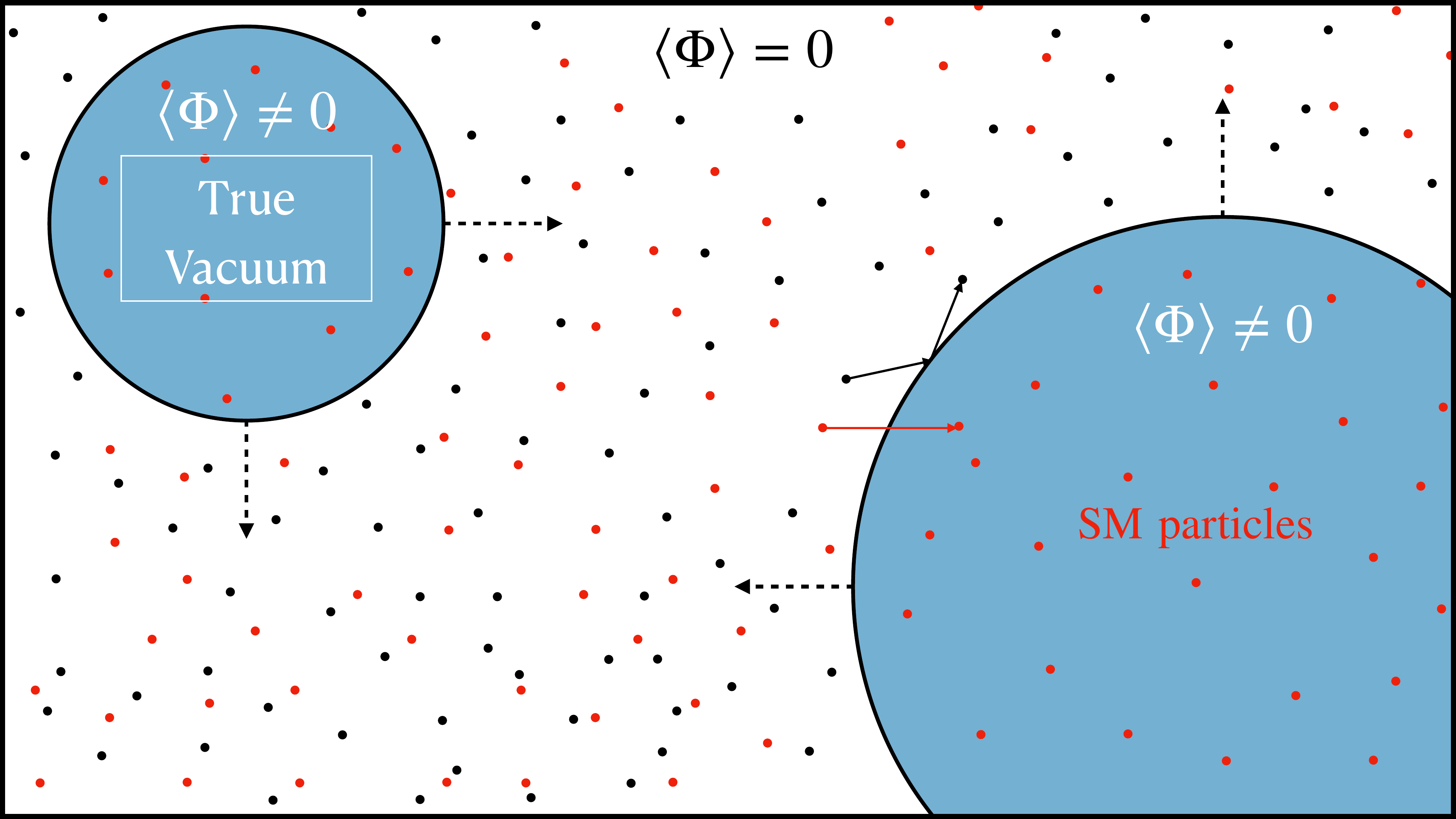}
    \includegraphics[width = 0.49\linewidth]{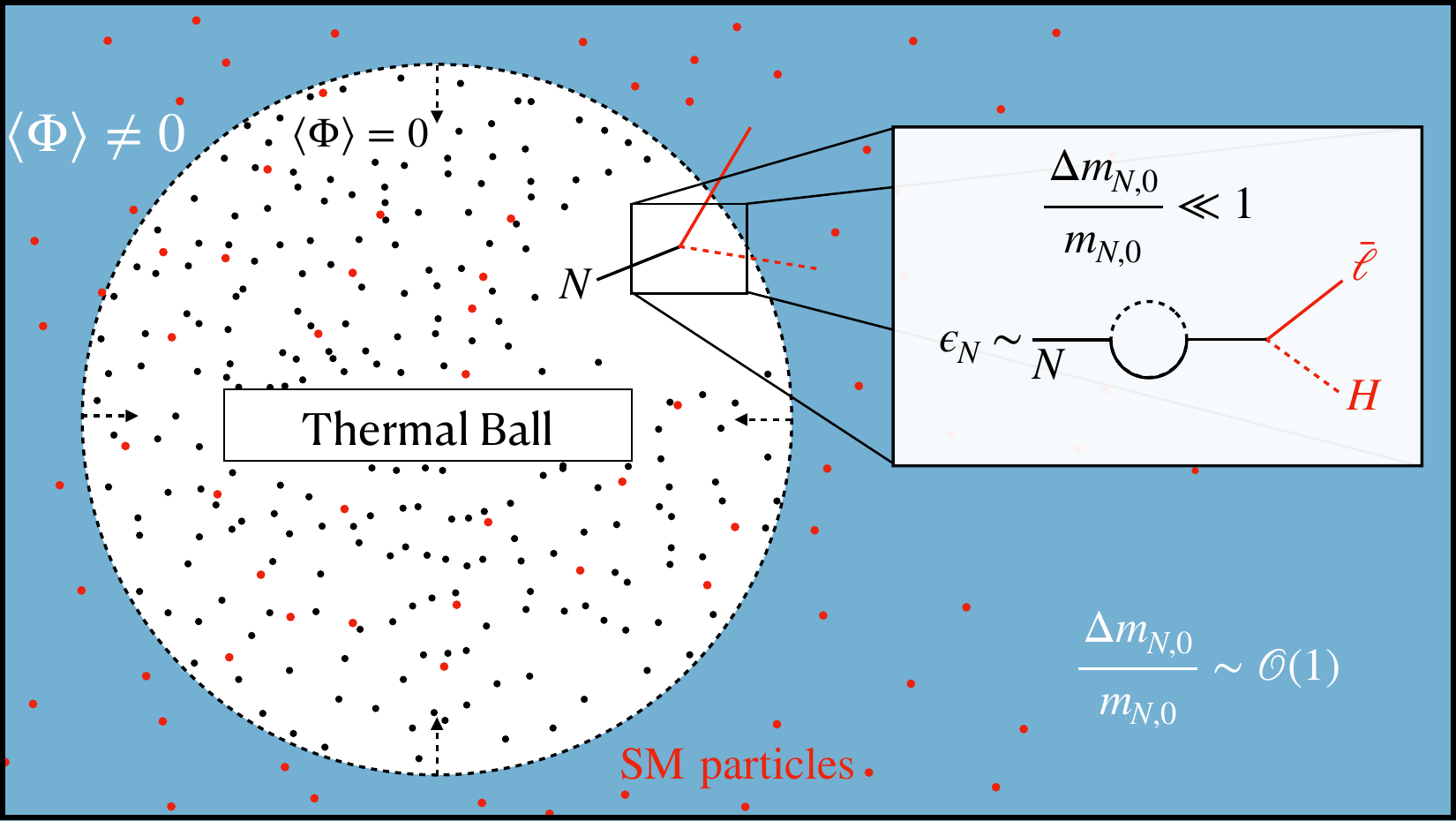}
    \caption{Illustration of phase separation baryogenesis. [Left] Nucleation of expanding bubbles due to FOPT. RHNs are unable to penetrate into the true vacuum where $\langle\Phi\rangle \neq 0$, while the SM particles can transit. [Right] As the bubbles expand, they eventually coalesce, resulting in confined remnants of the false vacuum where RHNs are trapped. Also shown are the relevant diagrams for the RHN decays that play a role in resonantly producing lepton asymmetry, where the RHN mass splitting $\Delta m_N$ is of order $\tilde\Gamma_N$.
    Resulting SM products of decaying RHNs freely penetrate bubble walls, with thermal balls evaporating.}
    \label{fig:cartoon}
\end{figure*}

{\it Mechanism and framework.}-- 
Different vacua phases can modify properties of decaying particles responsible for generating the baryon asymmetry, including their effective masses and couplings. This opens novel avenues for baryogenesis, as we demonstrate in the context of resonantly-enhanced production of a lepton asymmetry that is subsequently converted to a baryon asymmetry.

The scenario can be summarized as follows. In the context of cosmological first-order phase transitions (FOPTs),  RHNs can become trapped in the false vacuum when the processes that are relevant for leptogenesis are active.
Subsequently, pocket regions of the false vacuum result in the formation of thermal ball remnants that are stabilized by the thermal pressure from RHNs inside.
This holds the RHNs inside the thermal balls at a constant temperature $T_B$.
As we will show, in the false-vacuum phase  Majorana mass splitting of the RHNs $\Delta m_{N,0}$ satisfying $\Delta m_{N,0}/m_N \simeq 10^{-7}-10^{-3} $ can dynamically triggering resonant lepton asymmetry production. This contrasts with standard resonant leptogenesis, where the bare RHN mass splitting $\Delta m_{N,0}/m_N \simeq 10^{-12}-10^{-10}$ is typically required at low energy scales, demanding significant additional fine-tuning~\cite{Pilaftsis:2003gt}. On the other hand, in the true vacuum phase the RHNs can admit a hierarchical mass spectrum that would otherwise spoil the resonant enhancement of the asymmetry.
Hence, conditions on neutrinos for generating resonantly-enhanced lepton asymmetry are significantly relaxed compared to conventional models that require RHN mass degeneracy.

We consider a minimal model of leptogenesis where heavy RHNs $N$ 
couple to a new scalar $\Phi$. For generality, we do not restrict ourselves to specific ultraviolet (UV) completions.
We take a general Lagrangian that contains the interactions that generate the Dirac and Majorana mass terms
\begin{align}
    \mathcal{L} \supset \,&~h_{ij} \overline{N}_{i} \ell_{L_j}H - \frac{1}{2}(m_{N,0})_{ij} \overline{N}^c_{i} N_{j}  -  \frac{1}{2}(y_R)_{ij} \Phi \overline{N}^c_{i} N_{j}  \notag\\
    &+ V_{\rm eff}(\Phi,T) + {\rm h.c.} ,
\end{align}
where $\ell_{L_j}$ is the left-handed lepton doublet, $H$ is the Higgs doublet, $V_{\rm eff}(\Phi, T)$ is the effective finite temperature potential, $h_{ij}$ are the Dirac Higgs Yukawa couplings, $(y_R)_{ij}$ are the $\Phi$ Yukawa couplings, and $(m_N)_{ij}$ are Majorana mass contributions that we require to have small splittings $\Delta m_{N,0}$. Specific realizations including $V_{\rm eff}$ can be accounted for with additional terms. While our focus is on introducing a general mechanism, depending on specific model realizations additional observables can appear.
For example, if $\Phi$ couples to SM Higgs, new collider signatures can potentially constrain the parameters in $V_{\rm eff}$.

At some critical temperature $T_c$, $\Phi$ is taken to induce a first-order phase transition (FOPT). $\Phi$ obtains a vacuum expectation value (vev) $\langle\Phi\rangle = v$ and nucleates bubbles in the process. 
The FOPT can be described in terms of the parameters $\alpha_B = \Delta V/(\pi^2 g_B T_c^4/30)$ that characterizes its strength in the limit that $m_N \ll T_c$, and $\beta/H = (T d/dT (S_3/T))|_{T_*}$ with $T_* \simeq T_c$ that is related to its duration, where $S_3$ is the three-dimensional bounce action \cite{Witten:1980ez, Coleman:1977py}. 
In the massless RHN limit, the relativistic degrees of freedom in the thermal ball dark sector are $g_B \simeq 5.25$ for three generations of RHNs with the scalar $\Phi$ presumed too massive to be populated. 
Here, $\Delta V = V_{\rm eff} (\phi = 0) - V_{\rm eff}(\phi = v)$ is the difference in vacuum energy between the false and true vacua. These parameters depend on the exact details of the theory and control the transition dynamics. 
The FOPT leads to a RHN mass gain between the true and false vacua that is significant compared to thermal ball temperature $T_B$, with $y_R v/T_B \gg 1$, which confines the RHNs to the false vacuum pockets. Assuming thermal equilibrium inside thermal balls, the number density of semi-relativistic RHNs inside is
\begin{align}
    n_N^{}(T_B) &= \frac{g}{2\pi^2} T_B^3 \int_{0}^{\infty} dx~ x^2 
    \frac{1}{\exp(\sqrt{z^2 + x^2}) +1}\notag\\
    &= \frac{g}{2\pi^2} T_B^3 I(z),
\end{align} 
where $z = m_{N,T}/T_B$ and $x$ quantifies RHN momentum.

A schematic of the thermal ball formation, and the relevant dynamical RHN physical processes are shown in Fig.~\ref{fig:cartoon}.
For RHNs with mass above
$m_N > m_H + m_{\nu}$, decays into a Higgs boson of mass $m_H$  and a Standard Model neutrino can proceed via Yukawa interactions. When the RHN mass splitting is of order $\Delta m_{N,0} \sim \mathcal{O}(\Gamma_N)$, the quantity $\Delta m_{N,0}\, \tilde{\Gamma}_N/(\Delta m_{N,T}^2 + \tilde{\Gamma}_N^2/4)$ is resonantly enhanced, and the generated asymmetry can reach $\mathcal{O}(1)$. For the $z$ range we consider, we modify the decay rate by the relativistic $\gamma$ factor, $\tilde{\Gamma} = \Gamma \langle \gamma\rangle = \Gamma K_1(z)/K_2(z)$~\cite{Buchmuller:2004nz}, where $K_i$ is the modified Bessel function of the second kind. This factor takes into account velocity distribution of the RHNs in the thermal bath.
Importantly, the numerator of the asymmetry depends on the bare mass splitting $\Delta m_{N,0}$, while 
interactions with the SM bath at temperature $T_{\rm SM}$
induce thermally corrected mass difference in the denominator 
$\Delta m_{N, T} \sim {\rm Re}[(hh^{\dagger})_{ij}] T^2_{\rm SM}/m_{N,0}$, which are of the same order as the decay widths $\tilde{\Gamma}_N \sim  (hh^{\dagger})_{ii}m_{N,T} K_1(z)/K_2(z)$. These thermal corrections only slightly affect the location and height of the resonance peak by an $\mathcal{O}(1)$ factor, rather than spoiling it. Notably, due to the dynamics of the thermal balls, the mass splitting $\Delta m_{N,0}$ can be significantly larger than $\tilde{\Gamma}_N/2$ and still generate sufficient asymmetry.

Importantly, $\Phi$ can have non-degenerate couplings to the RHNs, which produce a hierarchical mass spectrum for the RHNs in the true vacuum. Naively, the non-degenerate $\Phi$ couplings might appear to give rise to non-degenerate thermal masses, therefore spoiling the small RHN mass splittings. However,
when the mass of $\Phi$ is significantly larger than the temperature in the false vacuum the thermal population of $\Phi$ in the false vacuum is exponentially suppressed. Hence, the thermal mass contributions to the RHN from the $\Phi$ will also be suppressed. Hierarchical values in the $y_R$ Yukawa coupling matrix can therefore generate hierarchical mass spectrum for the RHNs after $\Phi$ symmetry breaking in the true vacuum, while not destroying the degeneracy of masses needed in the false vacuum. Unique to our phase separation scenario, RHNs with nearly degenerate masses could decay resonantly in the false vacuum while having potentially observable hierarchical masses in the present Universe.

We consider the dominant source of thermal mass splitting to originate from the Dirac mass terms involving the Yukawa couplings $h_{ij}$. The RHNs have larger total Majorana masses $m_{N_i}$, which are quasi-degenerate in the false-vacuum phase. This setup can be naturally realized in a variety of models. For example, a symmetry can enforce a quasi-degeneracy of $m_{N_i}$ in the false vacuum. As the scalar $\Phi$ obtains a vev, 
non-degenerate Yukawa couplings $y_r$ can break such symmetry, generating a hierarchical RHN mass spectrum in the true vacuum.

For generality, we do not restrict ourselves to a specific scenario realization and note that other implementations are also possible. For example, our mechanism might also be realized when in the false vacuum the Majorana RHNs are relativistic. Then, additional relativistic effects beyond those considered in our analysis would need to be incorporated.  Nevertheless, the formation of thermal balls provides the necessary out of equilibrium condition and suppresses washout, potentially enabling successful leptogenesis even in the relativistic regime where standard mechanisms face significant challenges~\cite{Garbrecht:2019zaa}.

Our mechanism takes advantage of the broad parameter space complementary to that of earlier scenarios~\cite{Huang:2022vkf,Dichtl:2023xqd}
analyzing leptogenesis from RHNs squeezed into the true vacuum and decaying rapidly due to the large induced true vacuum RHN mass. In those scenarios such considerations led to the reduction in lepton asymmetry washout and an increase in the generated baryon asymmetry. However, this typically entailed requiring strong FOPTs with a supercooled phase. Instead, our mechanism can work with moderate and weak phase transitions that slow down due to friction force from the large mass gap and naturally form bound remnants. In our scenario, the presence of thermal balls enables efficient baryon asymmetry production even when the system is somewhat off resonance.  

{\it Thermal ball remnant formation.--}~  The FOPT facilitates dark sector particles in thermal bath becoming trapped inside macroscopic remnants of false vacuum. As $\Phi$ obtains a vev in true vacuum during FOPT, a significant mass gap of heavy RHNs $N$ is formed between true and false vacua and prevents them from penetrating the transition walls into the true vacuum due to insufficient kinetic energy. On the other hand, SM particles freely traverse to the true vacuum do to a lack of coupling to $\Phi$. Subsequently, RHNs
are compacted and confined between expanding true vacuum bubble walls resulting in formation of thermal balls, compact remnants with sufficiently high thermal pressure to balance the vacuum pressure differential.

The balance causes thermal balls to be held at a constant temperature~\cite{Lu:2022paj}
\begin{equation}
\label{balance temperature}
    T_{B}^{} = \left(3 \alpha_B\right)^{1/4} T_c~.
\end{equation}
The number density of the compact remnants depends on the FOPT parameters~\cite{Kawana:2021tde},
\begin{equation}
\label{eq:nT}
    n_B(T) = 1.31\times10^5 v_w^{-3}\left(\frac{\beta/H}{100}\right)^3 \frac{g_*(T)g_*(T_c)^{1/2}T^3 T_c^3}{M_{\rm pl}^3}~,
\end{equation}
where $M_{\rm pl} = 1.2 \times 10^{19}$ GeV is the Planck mass, $g_*$ accounts for the SM relativistic degrees of freedom, $v_w$ is the speed of the bubble walls.

The volume averaged radius of the thermal balls at formation is approximately
\begin{equation}
    R_B \simeq 1.28\left(\frac{1+\alpha_B}{1.16\alpha_B}\right)^{1/3} \frac{v_w}{\beta}~,
\end{equation}
derived in the massless RHN limit. The volume of the thermal balls at formation is $\mathcal{V}_B = 4\pi R_B^3/3$ and the volume fraction of the Universe occupied by thermal balls at formation is $f_V = n_B \mathcal{V}_B$. The energy trapped in  thermal balls is transmitted back into the SM plasma through SM particle production from RHN decay and inverse decay processes. This entropy injection can moderately reheat the SM. 

{\it Right-handed neutrino decays.}--
As the RHNs decay through $N \rightarrow H\ell$, the SM fields escape the thermal ball. Since the pressure balance is maintained inside, as the thermal ball cools it shrinks while releasing latent heat and compacting the plasma of dark sector fields so that its temperature remains constant. 
Correspondingly, the volume fraction of the Universe occupied by thermal balls $f_V$ decreases and eventually completely evaporate. The associated lepton asymmetry generation proceeds weighted by the remnant volume and the RHN decay rate. 
Due to the large $m_N$ in the true vacuum, washout processes are heavily suppressed. However, such processes may be active inside the remnants.

Immediately after the phase transition, the large thermal mass of the Higgs may prevent RHN decays $N \xrightarrow{} H \ell$. The Higgs thermal mass is quite large coming from from SM interactions, $m_{H,T}\sim 0.5 T_{\rm SM}$. 
The CP-violating RHN decays become kinematically allowed once the SM plasma redshifts to a temperature where the thermal mass satisfies $m_{H,T} < m_{N,T}$.
Most of the decays occur shortly after this threshold is crossed. For larger values of $m_{N,T}/T_B$, out of equilibrium decays may occur before thermal ball formation and generating additional lepton asymmetry while prematurely depleting the RHN population. To avoid this effect, we focus on the range $1\lesssim m_{N,0}/T_B \lesssim 10$ where such early decays are negligible. We begin our numerical analysis at the moment of thermal-ball formation.

We numerically solve a modified Boltzmann equation to track the RHN decays and associated processes, taking into account the thermal ball evolution, see Appendix for details. We consider RHN masses in the range of $1 \lesssim m_{N,0}/T_B \lesssim 10$. For lower masses $m_{N,0} \lesssim T_B$ a fully relativistic treatment~\cite{Garbrecht:2019zaa} including helicity-dependent effects would be required that is beyond the scope of this work.

We include the leading processes contributing to the lepton asymmetry and its washout. The RHNs are held at a constant temperature in the thermal ball while the SM temperature cools with the expansion of the Universe. While we focus here on RHN decay and inverse-decay processes, in other regimes where $m_N \lesssim T_B$, scatterings can also be relevant~\cite{Besak:2012qm,Mukaida:2012bz}.

{\it Baryon asymmetry.}-- Decays of the heavy RHNs produce lepton asymmetry that is subsequently converted to baryon asymmetry.
Near the resonance, the CP violation from RHN decays can be approximated as
\begin{align}
\label{eq:cpres}
    \epsilon_N =&~\frac{\sum_{\alpha}\left[\tilde{\Gamma}\left(N_{1} \rightarrow \ell_{\alpha} H\right)-\tilde{\Gamma}\left(N_{1} \rightarrow \bar{\ell}_{\alpha} \bar{H}\right)\right]}{\sum_{\alpha}\left[\tilde{\Gamma}\left(N_{1} \rightarrow \ell_{\alpha} H\right)+\tilde{\Gamma}\left(N_{1} \rightarrow \bar{\ell}_{\alpha} \bar{H}\right)\right]} \nonumber\\
    &~\sim \frac{1}{2}\frac{{\rm Im}\big(hh^{\dagger}\big)^2_{ij}}{(hh^{\dagger})_{ii} (hh^{\dagger})_{jj}} \frac{ \Delta m_{N,0} \,\tilde{\Gamma}_N}{(\Delta m_{N,T})^2 + \tilde{\Gamma}_N^2/4}~.
\end{align}
We note that the mass splittings in the numerator and denominator of Eq.~\eqref{eq:cpres} have different origins~\cite{Hohenegger:2014cpa}. In the numerator, $\Delta m_{N,0}$ comes from the Lagrangian bare-mass parameters, while the denominator $\Delta m_{N,T}$ arises from the propagators, which are modified by thermal corrections.

At the temperatures at which the lepton asymmetry is generated, the electroweak sphalerons are active and efficiently convert it to baryon asymmetry. As the sphalerons eventually freeze out, a resulting net baryon asymmetry originating from lepton asymmetry remains with the asymmetry conversion factor of $C_s \simeq 12/37$~\cite{Kuzmin:1985mm}.

\begin{figure}
    \centering  
    \includegraphics[width = 1.1\linewidth]{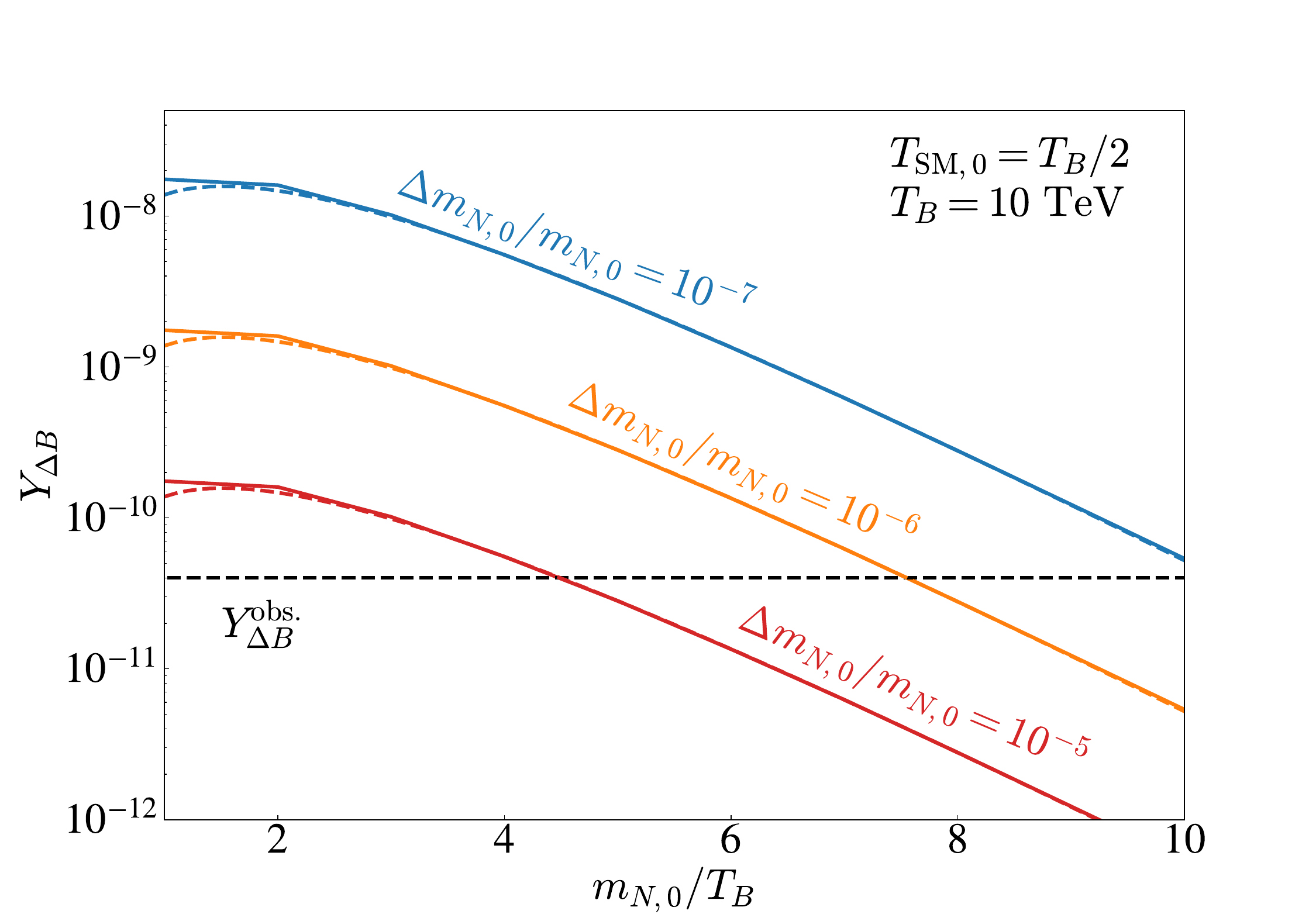}
    \caption{Generated baryon asymmetry as a function of $m_{N,0}/T_B$. Different possible suppression factors between the diagonal and off-diagonal Yukawa coupling matrix elements are shown (colored lines). The solid lines represent the numerical results, and the dashed lines are semi-analytical results from Eq.~\eqref{eq:analyticest}. The observed baryon asymmetry is indicated by the black dashed line. Characteristic parameter values for thermal ball temperature $T_B = 10$ TeV and initial SM temperature of $T_{\rm SM,\,0} = T_B/2$ are assumed.}
    \label{fig:asymmetry}
\end{figure}

In Fig.~\ref{fig:asymmetry} we display resulting baryon asymmetry from phase separation mechanism by numerically solving the Boltzmann equation, see Appendix for details.  
Intriguingly, we find that resonant lepton asymmetry production can involve CP asymmetries of order unity, thereby efficiently generating a baryon asymmetry that can even be much larger than the observed value of $\sim10^{-10}$. However, excess asymmetry can be additionally suppressed when there are additional CP conserving channels and new beyond SM fields. More so, the imaginary off-diagonal elements of the general Yukawa coupling matrix $(hh^{\dagger})_{ij}$ in Eq.~\eqref{eq:cpres} could be suppressed relative to the real diagonal elements in the denominator. We also illustrate effects of such possible suppression factors in Fig.~\ref{fig:asymmetry}.

We confirm good agreement of our numerical calculations with semi-analytical estimates of the resulting baryon asymmetry assuming RHNs decay without washout 
and that dark sector processes involving Yukawa couplings $y_R$ are sufficiently efficient for the RHNs and the scalar $\Phi$ to remain in equilibrium. Then, as the RHNs decay, the energy density of $\Phi$ is transferred to the RHNs and contributes to the generated lepton asymmetry. 

The RHN decay rate becomes sufficiently rapid when $\Gamma_N/H > 1$ is satisfied 
and is approximately instantaneous on cosmological timescales, for the parameters of interest.
The resulting asymmetry will be predominantly generated at an effective SM decay temperature $T_d \lesssim T_c$, the SM temperature that maximizes the CP asymmetry generation rate
$\epsilon_N \times d\log(f_V)/dt$. These decays transfer energy from the thermal balls to the surrounding SM plasma, which can increase the temperature. For the relevant parameter space we have confirmed that these effects are negligible so that the reheating temperature is approximately equal to the decay temperature $T_{\rm reh} \simeq T_d$ (see Appendix).

To obtain the total generated lepton asymmetry, we estimate the RHN number density $n_N$ at the epoch $T_d$. We consider for simplicity that all three species of RHNs have similar CP asymmetries and thus contribute equally. Additionally, 
we also include the scalar $\Phi$ and vacuum energy density from $\Delta V$ 
that would be converted into RHNs by 
efficient new physics processes before ultimately decaying. 
The average number density of RHNs over all space is $\langle n_{N} \rangle =n_N f_V$.

Each decay generates an average asymmetry $\epsilon_N$, which can be estimated from the resonant decay formula of Eq.~\eqref{eq:cpres} at $T_{\rm d}$. Neglecting washout, the resulting baryon asymmetry can then be estimated simply as 
\begin{align} 
\label{eq:analyticest}
     Y_{\Delta B} &~\simeq~ C_s \frac{\langle n_{N} \rangle (T_{B})}{s(T_{d})} 2 \epsilon_N \notag\\
     &~\simeq 1.4\times10^{-3} \left(\frac{T_B}{T_{\rm SM,0}}\right)^3 \epsilon_N(T_d) ~I\bigg(\frac{m_{N,T}}{T_B}\bigg) f_V~. 
\end{align}
The SM entropy is $s(T) = (2\pi^2/45)g_*(T) T^3$, with $g_*(T)$ the relativistic degrees of freedom in the SM plasma so that $g_B/g_*(T_{\rm SM,0}) = 5.25/106.75$ in our scenario. For comparison of the analytical estimates with our full numerical calculations we have computed $\epsilon_N$ at $T_d \simeq T_{\rm SM,0}$ and then, using Eq.~\eqref{eq:analyticest}, displayed the resulting analytic baryon asymmetry results (dotted lines) in Fig.~\ref{fig:asymmetry}. We find good agreement between our analytical estimates and the  numerical results.
Here, we have omitted additional $\mathcal{O}(1)$ effects on the generated asymmetry that can originate from the entropy released during RHN decays, which we discuss in Appendix. In Fig.~\ref{fig:asymmetry}, we take the benchmark of $T_{\rm SM,0} = T_B/2$. Considering even $\mathcal{O}(1)$ changes in $T_B/T_{\rm SM,0}$ can significantly enhance generated asymmetry, enabling $\Delta m_{N,0}/m_{N,0} \simeq 10^{-3}$
to yield the correct observed baryon asymmetry for
$T_B/T_{\rm SM,0} \simeq 5$.

In Fig.~\ref{fig:asymmetry}, we demonstrate that the correct baryon asymmetry can be achieved across the entire displayed parameter space. Moreover, additional possible non-CP-violating decay modes, depending on UV-complete models, can naturally mitigate asymmetry overproduction in regions where it occurs. This underscores the robustness of our mechanism, demonstrating that it allows for a broad viable parameter space and accommodates a variety of possible realizations.

Additional coincidence signatures can assist in probes of our scenario. Generically, FOPTs generate stochastic gravitational wave (GW) signals, which could be detected by future GW detectors. The spectrum of the GW signal depends on the details of the potential and model realization. Nevertheless, the peak GW frequency is expected to be $f_{\rm GW} \simeq O(1)\,\textrm{mHz}\,(\beta/H /100)\,(T /100~{\rm GeV})$, which could be detected by upcoming interferometers such as LISA \cite{LISACosmologyWorkingGroup:2022jok}, Einstein Telescope \cite{Branchesi:2023mws}, Cosmic Explorer \cite{Maggiore:2019uih}, Big Bang Observer \cite{Crowder:2005nr} and DECIGO \cite{Kawamura:2011zz}. 

In addition to generating the baryon asymmetry, the new scalar $\Phi$ may lead to potential collider signatures if its mass is near the TeV scale. For example, if 
$\Phi$ couples to the Higgs sector or other SM particles, it can in principle produce signals related to Higgs portal interactions, scalar resonances or exotic decays at colliders.

{\it Conclusions.}-- 
We have put forth a novel and general mechanism for generating the observed cosmological baryon asymmetry, which we call phase separation leptogenesis. This mechanism exploits the dramatically distinct behavior of decaying particles in different vacuum phases, where particle masses and couplings vary due to environmental effects. As a result, baryon asymmetry production can proceed efficiently without the constraints typically imposed by uniform vacuum cosmology.

We demonstrated a concrete realization of this mechanism in the context of resonant leptogenesis. In our scenario, RHNs trapped in false vacuum remnants can naturally satisfy the resonant condition without requiring significant fine-tuned mass degeneracies in the true vacuum. This allows to achieve low energy scale leptogenesis with hierarchical RHN masses, avoiding the   parameter space restrictions typically encountered for thermal leptogenesis at scales $\lesssim 10^9~{\rm GeV}$. Our mechanism thus relaxes the constraints on neutrino mass parameters compared to conventional approaches, reopening viable low energy scale leptogenesis parameter space.

Phase separation leptogenesis may provide a pathway to enhancing baryon asymmetry generation in the relativistic RHN regime, where standard leptogenesis mechanisms are typically ineffective due to washout. The false vacuum remnant thermal ball environment suppresses washout by construction, potentially enabling successful relativistic leptogenesis.  More generally, our framework complements existing true vacuum leptogenesis scenarios by allowing resonant RHN decays and asymmetry generation to occur within false vacuum remnants. This mechanism dynamically suppresses washout and eliminates the need for persistent fine-tuned RHN mass splittings in the late Universe. 

The phase separation mechanism is not restricted to just leptogenesis. In principle, other fundamental parameters such as coupling strengths, particle interactions, or decay rates could vary between true and false vacua, leading to a broad class of baryogenesis scenarios. These possibilities warrant further investigation. Additional experimental signatures, such as gravitational waves from the first-order phase transition, offer promising avenues to test and distinguish this framework.

~\newline
\textit{Acknowledgments.} We thank Kyohei Mukaida and Tim Tait for comments.  J.A. is supported by the NSF QLCI Award OMA - 2016244. J.A. thanks WPI-QUP center at KEK for hospitality, where part of this work was conducted. P.L. is supported by Grant Korea NRF2019R1C1C1010050 and KIAS Individual Grant 6G097701.  
V.T. acknowledges support by the World Premier International Research Center Initiative (WPI), MEXT, Japan and JSPS KAKENHI grant No. 23K13109.

\appendix

\section{Asymmetry from Right Handed Neutrino Decays and Inverse Decays}

We start with the general equation for the evolution of the distribution of leptons, $f_{\ell}$. Taking into account RHN decays and inverse decays, this is given by
\cite{Frossard:2012pc} 
\begin{align} \label{eq:lepdist}
    p^{\mu} \mathcal{D}_{\mu} f_{\ell} = &~ n_B \mathcal{V}_B\int \frac{d^3 q}{(2\pi)^3 2E_{q}} \frac{d^3 k}{(2\pi)^3 2E_{k}} (2\pi)^4 \delta^4(p-k-q) \notag\\
    &\times \frac{K_1(z)}{K_2(z)}\Big[|\mathcal{M}|^2_{N_i \rightarrow \ell H} (1 - f_{\ell})(1+f_H)f_N \\ &- |\mathcal{M}|^2_{\ell H \rightarrow N_i} f_{\ell}f_H (1-f_{N}) \Big]~,
\end{align}
\noindent where $p$, $k$, and $q$ are the momenta for RHNs $N$, leptons $\ell$, and Higgs $H$ respectively, $\mathcal{D}_{\mu}$ is the covariant derivative (with respect to the Friedmann-Robertson-Walker (FRW) metric), $f_i = f_i(E_i,T_i)$ is the $i$-th species particle distribution function, taking into account the thermal distributions of the different species of particles. Notably for the calculation here, the RHN temperature $T_B$ is different than the $\ell, H$ ($T_{\rm SM}$) temperature. Then, $|\mathcal{M}|^2_{N_i \leftrightarrow \ell H}$ is the matrix element for RHN decays and inverse decays. 
Here, the factor of the volume of the thermal balls $\mathcal{V}_B$ accounts for the fact that
thermal balls occupy only a fraction of the volume of the Universe $\mathcal{V}_H \sim 1/H^3$ defined by the Hubble parameter $H$,
with $n_B$ being the number density of thermal balls per Hubble volume.
As discussed in the main text, the processes are also modified by the relativistic Lorentz factor for the $z = m_{N,T}/T_B$ range we consider,  with $\langle \gamma \rangle = K_1(z)/K_2(z)$~\cite{Buchmuller:2004nz}.

An equivalent expression exists for the evolution of the anti-leptons, simply with the replacements $\ell \leftrightarrow \bar{\ell}$ and $H \leftrightarrow \bar{H}$. 
Subtracting the two distribution equations for leptons $f_l$ of Eq.~\eqref{eq:lepdist} and anti-leptons $f_{\bar{\ell}}$, we obtain
\begin{widetext}
\begin{align}
    p^{\mu} \mathcal{D}_{\mu} \bigg(f_{\ell} - f_{\bar{\ell}}\bigg) = & ~n_B \mathcal{V}_B\int \frac{d^3 q}{(2\pi)^3 2E_{q}} \frac{d^3 k}{(2\pi)^3 2E_{k}} (2\pi)^4 \delta^4(p-k-q)\frac{K_1(z)}{K_2(z)}\nonumber\\
    &\times\bigg\{|\mathcal{M}|^2_{N_i \rightarrow \ell H}[(1 - f_{\ell})(1+f_H)f_N + f_{\bar{\ell}}f_{\bar{H}} (1-f_{N})]- |\mathcal{M}|^2_{N_i \rightarrow \bar{\ell} \bar{H}} [(1 - f_{\bar{\ell}})(1+f_{\bar{H}})f_N + f_{\ell}f_H (1-f_{N})]\bigg\}~.
\end{align}
\end{widetext}
Here, we used identifications implied by
CPT charge, parity and time reversal symmetry invariance
\begin{align}
    |\mathcal{M}|^2_{N_i \rightarrow \ell H} &= |\mathcal{M}|^2_{\bar{\ell} \bar{H} \rightarrow N_i}\\
    |\mathcal{M}|^2_{N_i \rightarrow \bar{\ell} \bar{H}} &= |\mathcal{M}|^2_{\ell H \rightarrow N_i}~.
\end{align}

Using the definition of the asymmetry $\epsilon_N$
\begin{align}
    \epsilon_N = \frac{\tilde\Gamma_{N_i \rightarrow \ell H} - \tilde\Gamma_{N_i \rightarrow \bar{\ell} \bar{H}}}{\tilde\Gamma_{N_i \rightarrow \ell H} + \tilde\Gamma_{N_i \rightarrow \bar{\ell} \bar{H}}}
\end{align}
we can relate $|\mathcal{M}|^2_{N_i \rightarrow \ell H}$ and $|\mathcal{M}|^2_{N_i \rightarrow \bar{\ell} \bar{H}}$.
Rewriting this
\begin{widetext}
\begin{align}
    \epsilon_N \bigg(\tilde\Gamma_{N_i \rightarrow \ell H} + \tilde\Gamma_{N_i \rightarrow \bar{\ell} \bar{H}}\bigg) &= \tilde\Gamma_{N_i \rightarrow \ell H} - \tilde\Gamma_{N_i \rightarrow \bar{\ell} \bar{H}}\\
    \Rightarrow \epsilon_N \int d\Pi_H d\Pi_\ell \bigg(|\mathcal{M}|^2_{N_i \rightarrow \ell H} + |\mathcal{M}|^2_{N_i \rightarrow \bar{\ell} \bar{H}}\bigg) &= \int d\Pi_H d\Pi_\ell \bigg(|\mathcal{M}|^2_{N_i \rightarrow \ell H} - |\mathcal{M}|^2_{N_i \rightarrow \bar{\ell} \bar{H}}\bigg)
\end{align}
\end{widetext}
where the integrals shown are the decay width $\Gamma_{i\rightarrow j}$ for process $i \rightarrow j$, and the $d\Pi_{j}$ factors are denote the phase space elements for the final state, which must be integrated over to calculate the decay widths, such that $\Gamma_{i \rightarrow j} \sim \int d\Pi_{j} |{\cal M}|_{i\rightarrow j}^2$. Since $\epsilon_N$ is only a function of temperatures, masses, and couplings, it's a constant with respect to the phase space integrals. Therefore, we can make the identification,
\begin{widetext}
\begin{align}
     \int d\Pi_H d\Pi_\ell ~\epsilon_N \bigg(|\mathcal{M}|^2_{N_i \rightarrow \ell H} + |\mathcal{M}|^2_{N_i \rightarrow \bar{\ell} \bar{H}}\bigg) &= \int d\Pi_H d\Pi_\ell \bigg(|\mathcal{M}|^2_{N_i \rightarrow \ell H} - |\mathcal{M}|^2_{N_i \rightarrow \bar{\ell} \bar{H}}\bigg)\\
     \Rightarrow \epsilon_N \,(|\mathcal{M}|^2_{N_i \rightarrow \ell H} + |\mathcal{M}|^2_{N_i \rightarrow \bar{\ell} \bar{H}}) &= (|\mathcal{M}|^2_{N_i \rightarrow \ell H} - |\mathcal{M}|^2_{N_i \rightarrow \bar{\ell} \bar{H}})
\end{align}
\end{widetext}
We can therefore rewrite $|\mathcal{M}|^2_{N_i \rightarrow \bar{\ell} \bar{H}}$ as

\begin{align}
     |\mathcal{M}|^2_{N_i \rightarrow \bar{\ell} \bar{H}}
    &= \frac{1-\epsilon_N}{1+\epsilon_N}|\mathcal{M}|^2_{N_i \rightarrow \ell H}~.
\end{align}

\begin{figure*}
    \centering
    \includegraphics[width = 0.49\linewidth]{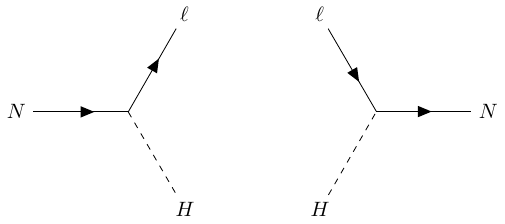}
    \includegraphics[width = 0.49\linewidth]{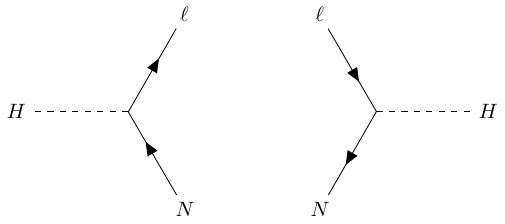}
    \caption{Tree-level diagrams describing relevant RHN processes that we consider for asymmetry generation.}
    \label{fig:enter-label}
\end{figure*}

Now, we can also calculate $|\mathcal{M}|^2_{N_i \rightarrow \ell H}$, which considering contributing tree-level classical processes of Fig.~\ref{fig:enter-label} is
\begin{align}
    |\mathcal{M}|^2_{N_i \rightarrow \ell H, \rm tree}  = 2 (hh^{\dagger})_{ij} (p\cdot k)~.
\end{align}

The total Boltzmann equation is obtained by integrating over the RHN phase space
\begin{widetext}
\begin{align}
    \frac{dn_{\ell- \bar{\ell}}}{dt} + 3 H n_{\ell- \bar{\ell}} \simeq&~ n_B \mathcal{V}_B \int \frac{d^3 p}{(2\pi)^3 2E_{p}}\frac{d^3 q}{(2\pi)^3 2E_{q}} \frac{d^3 k}{(2\pi)^3 2E_{k}} (2\pi)^4 \delta^4(p-k-q)\frac{K_1(z)}{K_2(z)}\notag\\
    &\times\bigg\{|\mathcal{M}|^2_{N_i \rightarrow \ell H}[(1 - f_{\ell})(1+f_H)f_N + f_{\bar{\ell}}f_{\bar{H}} (1-f_{N})]\nonumber\\
    &- \frac{1 - \epsilon_N}{1+\epsilon_N}|\mathcal{M}|^2_{N_i \rightarrow \ell H} [(1 - f_{\bar{\ell}})(1+f_{\bar{H}})f_N + f_{\ell}f_H (1-f_{N})]\bigg\}~.
\end{align}
\end{widetext}
We assume here that the distributions are nearly in equilibrium with $f_{i} \simeq f_{\bar{i}}$ \cite{Frossard:2012pc}. Additionally, to further simplify numerical integration it is convenient to expand $1 -  (1 - \epsilon)/(1+\epsilon) \simeq 2\epsilon$, which is a good approximation when $\epsilon \ll 1 $. This is valid outside of considering exact resonance, where $\epsilon_{\rm max} = 1/2$. Using both approximations we obtain
\begin{widetext}
\begin{align} \label{eq:boltful}
     \frac{dn_{\ell- \bar{\ell}}}{dt} + 3 H n_{\ell- \bar{\ell}} \simeq&~ ~n_B \mathcal{V}_B \int \frac{d^3 p}{(2\pi)^3 2E_{p}}\frac{d^3 q}{(2\pi)^3 2E_{q}} \frac{d^3 k}{(2\pi)^3 2E_{k}} (2\pi)^4 \delta^4(p-k-q) \notag\\
    &~\times  \frac{K_1(z)}{K_2(z)}2\,\epsilon_N(m_{N_i},h,T_{\rm SM})|\mathcal{M}|^2_{N_i \rightarrow \ell H}[(1 - f_{\ell}(E_k,T_{\rm SM}))(1+f_H(E_{q},T_{\rm SM}))f_N(E_p, T_{B}) \notag\\&~+ f_{\ell}(E_k,T_{\rm SM})f_H(E_{q},T_{\rm SM}) (1-f_{N}(E_p, T_{B}))]~.
\end{align}
\end{widetext}
Recasting the Boltzmann Eq.~\eqref{eq:boltful} in terms of comoving number density 
\begin{equation} \label{eq:comnumb}
 Y_{\ell - \bar{\ell}} \equiv \dfrac{n_{\ell - \bar{\ell}}}{s(T_{\rm SM})}
\end{equation}
with $x = m_{N,T}/T_{\rm SM}$. Denoting the right hand side of Eq.~\eqref{eq:comnumb} as $n_B \mathcal{V}_B \mathcal{I}_{\ell - \bar{\ell}}(T_{\rm SM})$, we obtain
\begin{align}
    \frac{dY_{\ell - \bar{\ell}}}{dx} = \frac{1}{H x s} n_B \mathcal{V}_B \mathcal{I}_{\ell - \bar{\ell}}(T_{\rm SM})~.
\end{align}
We then can integrate this as
\begin{align}
    \Delta Y_{\ell - \bar{\ell}} = \int dx \frac{1}{H x s} n_B \mathcal{V}_B \mathcal{I}_{\ell - \bar{\ell}}(T_{\rm SM})~.
\end{align}
However, integration requires accounting for thermal ball volume $\mathcal{V}_B$ scaling with the SM temperature $T_{\rm SM}$. Once the volume of thermal balls decreases to zero, the evolution of the Boltzmann equation is ceased.

To determine the evolution of the thermal ball size $\mathcal{V}_B$ with SM temperatures, we consider the Boltzmann equation for the distribution of RHNs within the thermal ball
\begin{align}
    &\frac{dn_N}{dt} + 3 \frac{\dot{R}_B}{R_B} n_N =~ - \int \frac{d^3 p}{(2\pi)^3 2E_{p}}\frac{d^3 q}{(2\pi)^3 2E_{q}} \frac{d^3 k}{(2\pi)^3 2E_{k}}\nonumber\\ &~\times(2\pi)^4 \delta^4(p-k-q) \{|\mathcal{M}|^2_{N_i \rightarrow \ell H} (1 - f_{\ell})(1+f_H)f_N \nonumber\\
    &~- |\mathcal{M}|^2_{\ell H \rightarrow N_i} f_{\ell}f_H (1-f_{N})+ (\ell H \leftrightarrow \bar{\ell} \bar{H})\}\frac{K_1(z)}{K_2(z)}~,
\end{align}
where $(\ell H \leftrightarrow \bar{\ell} \bar{H})$ denotes the same expressions for $\ell$ and $H$ but swapped with $\bar{\ell}$ and $\bar{H}$, and the dot represents a total time derivative $d/dt$. Note that there isn't a corresponding volume fraction term in front of the integral on the RHS. This is due to the Boltzmann equation keeping track of the distribution of RHN inside the thermal ball. Since the pressure balance enforces a constant number density of RHNs
\begin{equation}
n_N = n_0 = \frac{g}{2\pi^2} T_B^3 \int_{0}^{\infty} dx~ x^2 \bigg(\exp(\sqrt{z^2 + x^2}) +1\bigg)^{-1} ~,   
\end{equation}
with $z = m_{N,T}/T_B$
where the derivative term vanishes, and we are left with the second term on the LHS. Further identifying 
\begin{equation}
    3 \frac{\dot{R}_B}{R_B} = \frac{\dot{\mathcal{V}}_B}{\mathcal{V}_B}~,
\end{equation} 
we arrive at
\begin{align}
    \frac{1}{\mathcal{V}_B}\frac{d\mathcal{V}_B}{dt} &= - \frac{1}{n_0} \mathcal{I}_N(T_{\rm SM})~.
\end{align}

Here, $\mathcal{I}_N$ represents the entire phase space integral the contributes to the depletion of RHNs inside the thermal ball, $\mathcal{V}_B$ is the volume of the thermal ball, and $N_N = n_N \mathcal{V}_B$ is the total number of RHNs in the thermal ball. 
We assume a radiation dominated universe, which leads to $d/dt = -H(T_{\rm SM}) T_{\rm SM} d/dT_{\rm SM}$. Integrating, we obtain
\begin{align}
    \mathcal{V}_B(T_{\rm SM}) = \mathcal{V}_0 \exp\bigg(-\int^{T_{{N}}}_{T_{\rm SM}} \frac{\mathcal{I}_N(T^{\prime}_{\rm SM})}{n_0 T_{\rm SM}^{\prime 3}} dT^{\prime}_{\rm SM}\bigg)~.
\end{align}

The combined equation for the asymmetry is then,
\begin{align}
    \Delta Y_{\ell - \bar{\ell}} =& \int \frac{n_B \mathcal{V}_0}{H(T_{\rm SM}) T_{\rm SM} s(T_{\rm SM})}  \mathcal{I}_{\ell - \bar{\ell}}(T_{\rm SM}) \nonumber\\
    &\times\exp\bigg(-\int^{T_{{N}}}_{T_{\rm SM}} \frac{\mathcal{I}_N(T^{\prime}_{\rm SM})}{n_0 T_{\rm SM}^{\prime 3}} dT^{\prime}_{\rm SM}\bigg)  dT_{\rm SM}~.
\end{align}

\section{Phase Space Integrals}

There are two phase space integrals that need to be evaluated. One for the evolution of the lepton asymmetry generated by RHN interactions $\mathcal{I}_{\ell -\bar\ell}$, and the other for tracking the population of the RHNs themselves $\mathcal{I}_N$. 

\subsection{$\cal{I}_{\ell-\bar\ell}$}

The integral that corresponds to the interaction terms in the Boltzmann equation that generate the lepton asymmetry is given by

\begin{align}
   I_{\ell - \bar\ell} =  &\int \frac{d^3 p}{(2\pi)^3 2E_{p}}\frac{d^3 q}{(2\pi)^3 2E_{q}} \frac{d^3 k}{(2\pi)^3 2E_{k}} (2\pi)^4 \delta^4(p-k-q) \notag\\
    &~\times  \frac{K_1(z)}{K_2(z)}2\,\epsilon_N(m_{N_i},h,T_{\rm SM})|\mathcal{M}|^2_{N_i \rightarrow \ell H}\notag\\
    &~\times [(1 - f_{\ell}(E_k,T_{\rm SM}))(1+f_H(E_{q},T_{\rm SM}))f_N(E_p, T_{B}) \notag\\&~+ f_{\ell}(E_k,T_{\rm SM})f_H(E_{q},T_{\rm SM}) (1-f_{N}(E_p, T_{B}))]~
\end{align}

Although the final integral will need to be numerically integrated, we can perform the $q$ and $k$ integrals using the delta function, taking care of the $q$ integral first with $\delta^3(
\bf{p}-\bf{q}-\bf{k})$. Then, the remaining delta function $\delta(E_p - E_k - E_q)$ can be rewritten in terms of a delta function in the momentum $k$. The delta function can be changed from energies to momenta via

\begin{widetext}
\begin{align}
    \delta(E_p-E_k-E_{p-k}) = \delta(|\textbf{k}| - k_0) \bigg( 1 + \frac{\frac{m_N^2 - m_H^2}{2(E_p - p\cos\theta)}- p\cos\theta}{\sqrt{m_H^2 + p^2 + \frac{(m_H^2 - m_N^2)^2}{4(E_p - p\cos\theta)^2} + \frac{(m_H^2 - m_N^2)p\cos\theta}{E_p - p\cos\theta}}} \bigg)^{-1}
\end{align}
\end{widetext}

\noindent with $k_0 = \frac{1}{2}(m_N^2 - m_H^2)/(E_p - p\cos\theta)$. We can also write the phase space element in spherical coordinates, and integrate over angle $\varphi$, $d^3k \rightarrow 2\pi\, k^2 \, dk\, d\cos\theta$ and perform the $k$ integral, setting $k = k_0$. Then, using $p = \sqrt{E_p^2 - m_N^2}$,  $E_{k_0} = k_0$, and $E_{p-k_0} = \sqrt{(p-k_0)^2+m_H^2}$ gives the integral,

\begin{widetext}
\begin{align}
    \cal{I}_{\ell - \bar\ell} =  & (hh^{\dagger})_{ij}\frac{K_1(z)}{K_2(z)}\frac{\,\epsilon_N(m_{N_i},h,T_{\rm SM})}{4\pi^2}\int  dE_p \, d\cos\theta  \frac{\sqrt{E_p^2 - m_N^2}~ k_0^2}{E_{p-k_0}+ \frac{m_N^2 - m_H^2}{2(E_p - \sqrt{E_p^2 - m_N^2}\cos\theta}- \sqrt{E_p^2 - m_N^2}\cos\theta)} \notag\\
    &~\times  (E_p - \sqrt{E_p^2 - m_N^2}\cos\theta) \,[(1 - f_{\ell}(E_{k_0},T_{\rm SM}))(1+f_H(E_{p-k_0},T_{\rm SM}))f_N(E_p, T_{B}) \notag\\&~+ f_{\ell}(E_{k_0},T_{\rm SM})f_H(E_{p-k_0},T_{\rm SM}) (1-f_{N}(E_p, T_{B}))]~
\end{align}
\end{widetext}

\noindent We finally numerically integrate over $E_p$ and $\cos{\theta}$.

\subsection{$\cal{I}_N$}

Now, we similarly work on the phase space integral for the RHN within the thermal ball,

\begin{align}
    I_N &=~ - \int \frac{d^3 p}{(2\pi)^3 2E_{p}}\frac{d^3 q}{(2\pi)^3 2E_{q}} \frac{d^3 k}{(2\pi)^3 2E_{k}}\nonumber\\ &~\times(2\pi)^4 \delta^4(p-k-q) \{|\mathcal{M}|^2_{N_i \rightarrow \ell H} (1 - f_{\ell})(1+f_H)f_N \nonumber\\
    &~- |\mathcal{M}|^2_{\ell H \rightarrow N_i} f_{\ell}f_H (1-f_{N})\notag\\
    &~+ |\mathcal{M}|^2_{N_i \rightarrow \bar\ell \bar H} (1 - f_{\bar\ell})(1+f_{\bar H})f_N \nonumber\\
    &~- |\mathcal{M}|^2_{\bar\ell \bar H \rightarrow N_i} f_{\bar\ell}f_{\bar H} (1-f_{N})\}\frac{K_1(z)}{K_2(z)}~,
\end{align}

Once again using the CPT reversal symmetry,

\begin{align}
    |\mathcal{M}|^2_{N_i \rightarrow \ell H} &= |\mathcal{M}|^2_{\bar{\ell} \bar{H} \rightarrow N_i}\\
    |\mathcal{M}|^2_{N_i \rightarrow \bar{\ell} \bar{H}} &= |\mathcal{M}|^2_{\ell H \rightarrow N_i}~,
\end{align}
and assuming $f_{\bar\ell} = f_{\ell}$ and $f_{\bar H} = f_{H}$ are nearly in equilibrium for the purposes of the integration, and $|\mathcal{M}|^2_{N_i \rightarrow \bar\ell \bar H} = \frac{1-\epsilon}{1+\epsilon}|\mathcal{M}|^2_{N_i \rightarrow \ell H}$, the expression becomes 

\begin{widetext}
\begin{align}
    I_N &=~ - \int \frac{d^3 p}{(2\pi)^3 2E_{p}}\frac{d^3 q}{(2\pi)^3 2E_{q}} \frac{d^3 k}{(2\pi)^3 2E_{k}}(2\pi)^4 \delta^4(p-k-q) \nonumber\\ &~\times \{|\mathcal{M}|^2_{N_i \rightarrow \ell H} ((1 - f_{\ell})(1+f_H)f_N - f_{\ell}f_{H} (1-f_{N})) + \frac{1-\epsilon}{1+\epsilon}|\mathcal{M}|^2_{N_i \rightarrow \ell H} ((1 - f_{\ell})(1+f_{H})f_N - f_{\ell}f_H (1-f_{N}))\}\frac{K_1(z)}{K_2(z)}\nonumber\\
    &\approx~ - \int \frac{d^3 p}{(2\pi)^3 2E_{p}}\frac{d^3 q}{(2\pi)^3 2E_{q}} \frac{d^3 k}{(2\pi)^3 2E_{k}}(2\pi)^4 \delta^4(p-k-q) (2 - 2\epsilon)\{|\mathcal{M}|^2_{N_i \rightarrow \ell H} ((1 - f_{\ell})(1+f_H)f_N - f_{\ell}f_{H} (1-f_{N})) \}\frac{K_1(z)}{K_2(z)}
\end{align}
\end{widetext}

This is very similar to the asymmetry integral, so we can use many of the same techniques.

\begin{widetext}
\begin{align}
    I_{N} =  & -(hh^{\dagger})_{ij}\frac{K_1(z)}{K_2(z)}\frac{\,1-\epsilon_N(m_{N_i},h,T_{\rm SM})}{4\pi^2}\int  dE_p \, d\cos\theta  \frac{\sqrt{E_p^2 - m_N^2}~ k_0^2}{E_{p-k_0}+ \frac{m_N^2 - m_H^2}{2(E_p - \sqrt{E_p^2 - m_N^2}\cos\theta}- \sqrt{E_p^2 - m_N^2}\cos\theta)} \notag\\
    &~\times  (E_p - \sqrt{E_p^2 - m_N^2}\cos\theta) \,[(1 - f_{\ell}(E_{k_0},T_{\rm SM}))(1+f_H(E_{p-k_0},T_{\rm SM}))f_N(E_p, T_{B}) \notag\\&~- f_{\ell}(E_{k_0},T_{\rm SM})f_H(E_{p-k_0},T_{\rm SM}) (1-f_{N}(E_p, T_{B}))]~
\end{align}
\end{widetext}

This integral is also performed numerically. Notice that aside from numerical factors, the biggest difference is the relative signs between the two distribution function terms.

\section{Right-handed Neutrino Decay Temperature}

To semi-analytically estimate the asymmetry as discussed around Eq.~\ref{eq:analyticest}, we consider RHN decays that we approximate as being nearly instantaneous. It is then necessary to determine RHN decay temperature $T_d$, which we do as follows. 

The volume of thermal balls decreases as RHNs decay, and we choose the $T_{\rm SM}$ that corresponds to average asymmetry generation 
\begin{align}
    \frac{dn_{N}}{dt} + 3 \frac{\dot R}{R} n_N = -\Gamma_N (n_N-n^{\rm eq}_{N}(T_{\rm SM}))~.
\end{align}
As established above, the RHN density $n_N$ is constant, since $T_B$ is constant. We can therefore turn this into a differential equation for the volume of thermal balls. That is
\begin{align}
    3 \frac{\dot R}{R} = -\Gamma_N (T_{\rm SM}) \bigg(1-\frac{n^{\rm eq}_{N}(T_{\rm SM})}{n_N}\bigg)~.
\end{align}
Separating variables, we can relate the time $t$ to SM temperature $T_{\rm SM}$, assuming a radiation dominated universe, via $dt =  dT_{\rm SM}/(H(T_{\rm SM})T_{\rm SM})$. Integrating, we obtain
\begin{align}
    \log\bigg(\frac{\mathcal{V}_B}{\mathcal{V}_0}\bigg) = \int \frac{dT^{\prime}_{\rm SM}}{H T^{\prime}_{\rm SM}} \Gamma_N \bigg(1 - \frac{n^{\rm eq}_N(T_{\rm SM}^{\prime})}{n_N}\bigg)~.
\end{align}

Then, the relevant decay time is when most of the asymmetry is generated. We can characterize this as
\begin{align}
    \frac{d}{dT_{\rm SM}}\bigg(\epsilon_N \frac{d \log{\mathcal{V}_B}}{dt}\bigg)\bigg|_{T_d} = 0
\end{align}
where $T_d$ is defined as the temperature where the asymmetry generated during the process is maximized.

\section{Analytic Asymmetry Approximation with Entropy Release}

We can evaluate the impact of the energy and entropy transfer from the thermal balls to the SM sector during decays as follows. These factors are omitted in the main text for the parameter range $f_B \lesssim 1$.  The ratio of the energy in thermal balls relative to the SM plasma at the decay temperature $T_d$ is
\begin{align}
\begin{split}
\label{eq:fb}
    f_{\rm B}  
    \xrightarrow{z\ll1}  \frac{4}{3}\frac{g_B}{g(T_{\rm d})}\frac{T_c}{T_{\rm d}}~, 
\end{split}
\end{align}
which follows from energy conservation with $n_B$ and $E_B$ the number density and total energy of thermal balls. The decay of the RHNs releases energy into the SM plasma, with the temperature after this instantaneous decay being
\begin{equation}
\label{eq:treh}
    T_{\rm reh} = T_{\rm d}(1+f_{B})^{1/4}~.
\end{equation}
With the increased temperature of the SM plasma, the entropy $s$ is increased by a factor of $(1+f_B)^{3/4}$. Taking this effect into consideration and keeping track of $\alpha_B$, Eq.~\eqref{eq:analyticest} should be modified to
\begin{align}
\begin{split}
     Y_{\Delta B}
     \simeq & ~1.4\times10^{-3} \left(\frac{T_B}{T_d}\right)^3 (1+f_{\rm B})^{-3/4}\left(\frac{1+\alpha_B}{1.16\alpha_B}\right)  \epsilon_N(T_d)\notag\\
     ~&\times I\bigg(\frac{m_{N,T}}{T_B}\bigg) f_V ~.
\end{split}
\end{align}
In very strong FOPTs, the entropy released during decays could lead to a dilution of the lepton asymmetry. However, for our parameters of interest, these additional considerations contribute just an $\mathcal{O}(1)$ factor.

\bibliography{references}

\end{document}